\documentclass[aps,preprint,epsfig,rotate]{revtex4}
\usepackage{graphicx}
\usepackage{bm}
\usepackage{epsfig}

\begin{document}
\title{On the symmetry definitions for the constraint dynamical systems} 
       
\author{Alexei M. Frolov}
\email[E--mail address: ]{afrolov@uwo.ca}

\affiliation{Department of Applied Mathematics \\
 University of Western Ontario, London, Ontario N6H 5B7, Canada}

\date{\today}

\begin{abstract}

The problem of proper symmetry definition for constraint dynamical systems with Hamiltonians is considered. Finally, we choose a definition 
of symmetry which agrees with the analogous definition used for the non-constraint dynamical systems with Hamiltonians. Our symmetry definition 
allows one to consider the whole spectrum of the Hamiltonian without splitting it into a few different parts.

\end{abstract}

\maketitle
\newpage

Let us consider dynamical quantum systems each of which has the Hamiltonian operator $\hat{H}$. The corresponding Schr\"{o}dinger equation for such a system takes 
the form
\begin{equation}
  \hat{H} \Psi = E \Psi \; \; \; , \; \; or \; \; \;  \hat{H} \Psi = \imath \hbar \frac{\partial}{\partial t} \Psi \; \; \; \label{equ1}
\end{equation}
where $\Psi$ is the wave function, while $\hbar$ is the reduced Planck constant ($\hbar = \frac{h}{2 \pi}$) and $\imath$ is the imaginary units. The parameter $E$ in 
Eq.(\ref{equ1}) is the total energy of the quantum system. The first equation in Eq.(\ref{equ1}) is the time-independent Schr\"{o}dinger equation, while the second 
equation is the time-dependent Schr\"{o}dinger equation. In reality, many quantum systems have additional symmetry, e.g., geometrical, dynamical and/or `hidden' 
symmetry. Briefly, such a symmetry means invariance of the Schr\"{o}dinger equation for some (closed) group of physical, finite transformations. The `conditions of 
invariance' can be re-written in terms of infinitesimal (or contact) transformations which form a closed algebraic structure in terms of commutation relations between
each pair of the corresponding generators. In general, each group of such transformations has finite number of generators of infinitesimal transformations which are
represented as self-adjoint operators. These generators are designated below by the notation $A_{i}$, where $i = 1, \ldots, N$. They form a closed structure which is 
called the algebra Lie of infinitesimal transformations. The corresponding group of the finite (or physical) transformations is uniformly reconstructed, if its algebra 
Lie is known. The commutator between each pair of generators of the Lie algebra plays a fundamental role in the whole theory of symmetry. It was shown by Sophus Lie 
(see \cite{Lie} and references therein) that such a commutator is always written in the form $[ A_{i}, A_{j} ] = \sum_{k} C^{k}_{ij} A_{k}$, where $C^{k}_{ij}$ are the 
group constants, or structural constants of the corresponding Lie algebra. As follows directly from the definition of the commutator the group constants $C^{k}_{ij}$ 
must be antisymmetric upon the both $i$ and $j$ indexes, i.e. $C^{k}_{ij} = -C^{k}_{ji}$. From here one finds that, e.g., $[ A_{i}, A_{j}] = - [ A_{j}, A_{i}]$ and 
$[ A_{i}, A_{i} ] = 0$. All fundamental facts about groups, their Lie algebras, etc can be found, e.g., in \cite{Pontr}, \cite{Eisen}. Here we do not want to repeat 
these definitions and discuss properties which follow from such definitions. Instead, we want to consider some possible definitions of physical symmetry.

One of the first definitions of symmetry in Quantum Mechanics was formulated in terms of the generators $A_i$ of the Lie algebra as the set of conditions $[ A_{i}, 
\hat{H}] = 0$, where $i = 1, \ldots, N$ and $N$ is the total number of generators in the Lie algebra. However, in the middle of 1960's such a definition was found to 
be quite restrictive in actual applications. In particular, that definition was applicable only to the energy levels (or states) of the system which have the same 
energy. It was not possible to move up and/or down along the spectra of states. Since then another extended symmetry definition has been proposed and applied. The 
extended definition can be formulated in the two different (but equivalent) forms: (a) $[ A_{i}, \hat{H}] \Psi = 0$ for $i = 1, \ldots, N$, and (b) $A_{i} \hat{H} 
\Psi = \hat{H} A_{i} \Psi $ for $i = 1, \ldots, N$, where $\Psi$ is the solution of the Schr\"{o}dinger equation, i.e. $(\hat{H} - E) \Psi = 0$. The first definition 
means that all generators of the Lie algebra commute with the Hamiltonian on solutions of the Schr\"{o}dinger equation. The second definition means that the generators 
$A_{i}$ of the Lie algebra transforms one solution of the Schr\"{o}dinger equation into another solution of the same Schr\"{o}dinger equation, i.e. if $\Psi$ is a 
solution of the Schr\"{o}dinger equation, then the functions $\Phi_{i} = A_i \Psi$ for $i = 1, \ldots, N$ are also its solutions. Such a definition represents the 
`dynamical' symmetry, i.e. the symmetry which is only important for the actual (dynamical) motion of the system, or motion which agrees with the dynamics of quantum 
system. An obvious difference with the old-fashioned definition of symmetry is clear. Below, we shall use only the dynamical symmetry definition. 

Now, we need to make another step forward and discuss a few possible definitions of symmetry for constrained Hamiltonian systems, i.e. for quantum systems which have 
Hamiltonians and a number of constraints. In this study, we assume that all constraints are the first class constraints. An important example of the constrained 
Hamiltonian systems is the free electromagnetic field. At the end of 1920's the quantization of the free electromagnetic field was a serious problem, since it was clear 
that the two gauge conditions $\frac{\partial \phi}{\partial t} = 0$ and $div {\bf A} = 0$ cannot be imposed on the components of the four-potential $(\phi, {\bf A})$ 
of this field. In reality, it did lead to very serious contradictions in the whole quantization procedure for the four-vector $(\phi, {\bf A})$. Fermi \cite{Fer} proposed 
an effective approach which allows one to solve all such troubles at once. Fermi \cite{Fer} assumed that the conditions $\frac{\partial \phi}{\partial t} = 0$ and $div 
{\bf A} = 0$ for the components of 4-potential must be replaced by the corresponding conditions for the field wave function $\Psi$, i.e. $\Bigr(\frac{\partial 
\phi}{\partial t}\Bigr) \mid \Psi \rangle = 0$ and $(div {\bf A}) \mid \Psi \rangle = 0$, where the notation $\mid \Psi \rangle$ stands for the wave function $\Psi$, i.e. 
$\Psi = \mid \Psi \rangle$. Moreover, only such wave functions $\Psi$ must be considered at the following steps of the procedure. Dirac immediately realized that we are 
dealing with the new Hamiltonian mechanics, which leads to the new type of motion in the Hamiltonian systems with constraints. Later such systems were called the 
constraint dynamical systems. In reality, it took almost 20 years for Dirac to develop the closed theory of the Hamiltonian systems with constraints \cite{Dir1}. Below 
we discuss only a restricted version of this theory (analysis of more general cases can be found, e.g., in \cite{Tyut}). 

The total Hamiltonain $\hat{H}_{tot}$ of an arbitrary quantum system with constraints is represented as the sum of its dynamical part $\hat{H}_{d}$ and constraint part 
$\hat{H}_c$, i.e. $\hat{H}_{tot} = \hat{H}_{d} + \hat{H}_{c}$, where $\hat{H}_{c} \Psi = 0$. Let us assume that we are dealing with the system which has $N_p$ primary 
constraints $\hat{p}_{i}$, $N_s$ secondary constraints $\hat{s}_{j}$ and $N_t$ tertiary constraints $\hat{t}_{k}$, where $N_p \ge N_s \ge N_t$. The constraint part of 
the Hamiltonian $\hat{H}_{tot}$ is represented as a linear function of the primary, secondary and tertiary constraints, i.e.     
\begin{equation}
 \hat{H}_{tot} = \hat{H}_{d} + \hat{H}_{c} = \hat{H}_{d} + \sum^{N_p}_{i=1} v_i \hat{p}_{i} + \sum^{N_s}_{j=1} u_j \hat{s}_{j} + \sum^{N_t}_{k=1} w_k \hat{t}_{k} 
 \label{equ2}
\end{equation}
According to the definition of the primary, secondary and tertiary constraints \cite{Dir2} we can write
\begin{eqnarray}
  [ \hat{p}_i, \hat{H}_d ] = \sum^{N_s}_{l=1} a_{il} \hat{s}_{l} + \sum^{N_t}_{m=1} b_{im} \hat{t}_{m} \; \; \; ,  \; \; \; 
  [ \hat{s}_j, \hat{H}_d ] = \sum^{N_t}_{l=1} c_{jq} \hat{t}_{q} \; \; \; ,  \; \; \;  [ \hat{t}_k, \hat{H}_d ] = 0 \; \; \; \label{equ3}
\end{eqnarray}
where some of the numerical coefficients $a_{il}, b_{im}$ and $c_{jq}$ can be equal zero identically. Note that the numerical coefficients $a_{il}, b_{im}, c_{jq}$ in 
Eq.(\ref{equ3}) and $v_i, u_j, w_k$ in Eq.(\ref{equ2}) are the field depended values, while the group constants $C^{k}_{ij}$ defined above cannot depend upon these values,
i.e. they are truly constants. It follows directly from Eq.(\ref{equ2}) and definitions of the constraints that $(\hat{H}_{tot} - \hat{H}_{d}) \Psi = 0$. Therefore, we can 
write
\begin{eqnarray}
   ( \hat{H}_{tot} - E) \Psi = ( \hat{H}_{d} - E ) \Psi = 0 \; \; \; \label{equ4}
\end{eqnarray} 

At this point we need to propose an accurate and workable definition of the physical symmetry which can be applied for constraint dynamical systems with Hamiltonians. First 
of all, it is clear that the dynamical part of the Hamiltonian $H_{d}$ must commute with all generators $A_{i}$ of the contact Lie algebra on solutions of the Schr\"{o}dinger 
equation, i.e. $[A_{i}, H_{d}] \Psi = 0$, or $A_{i} H_{d} \Psi = H_{d} A_{i} \Psi$. In other words, if $\Psi$ is the solution of the Schr\"{o}dinger equation with the 
Hamiltonian $H_{d}$, then  $A_{i} \Psi$ (for $i = 1, \ldots, N$) are also solutions of the same equation. Briefly, this means that if $(H_d - E) \Psi = 0$, then we also 
have $(H_d - E) (A_i \Psi) = 0$ for $i = 1, \ldots, N$. The second part of this definition must contain information which allows one to determine the commutation relations 
between generators of the Lie algebra and operators which represent constraints. Formally, we can write for the primary constraints $A_{\alpha} \hat{p}_{i} \Psi = \hat{p}_{i} 
(A_{\alpha} \Psi)$. Since $\hat{p}_{i} \Psi = 0$, then we have $\hat{p}_{i} (A_{\alpha} \Psi) = 0$, i.e. we have new primary constraints which are defined on the set of 
functions $\phi_{\alpha} = A_{\alpha} \Psi$, where $\alpha = 1, \ldots, N_p$. Analogously, for the secondary and tertiary constraints, i.e. $A_{\alpha} \hat{s}_{j} \Psi = 
\hat{s}_{j} (A_{\alpha} \Psi) = 0$ and $A_{\alpha} \hat{t}_{j} \Psi = \hat{t}_{j} (A_{\alpha} \Psi) = 0$. It is important to note that the total numbers of primary, secondary 
and tertiary constraints cannot increase during applications of the symmetry operators. Furthermore, application of the symmetry operation does not mix constraints, i.e. after
application of the symmetry operations all primary constraints are represented as the linear combinations of the primary constraints only. The same statement is true for all 
secondary and tertiary constraints. After an extensive analysis it became clear that alternative definitions of symmetry which allow to mix constraints lead to some fundamental
changes in the dynamics of quantum system. Therefore, such definitions cannot be accepted.       

It should be mentioned that there are some additional relations between constraints known for the dynamical systems. For instance, in our paper \cite{Our1} on the free gravity 
fields it was shown that this system has four primary $\hat{p}_i$ and four secondary $\hat{s}_i$ constraints (no tertiary constraints have been found). It was shown in 
\cite{Our1} that the Poisson between the four primary constraints and four secondary constraints equals to the product of the metric tensor $g^{\mu\nu}$ (with the additional 
coefficient $-\frac12$ and secondary constraint with two temporal indexes (or (00) constraint) (for more details, see \cite{Our1}). Very likely, any non-zero Poisson bracket
between different constraints for one dynamical systems must be represented as a linear combination of other constraints and operator $(H_d - E)$. The coefficients of such 
linear combination are some filed-dependent functions, i.e. they are not constants. In general, this statement has never been proved. However, in those cases when the wave 
functions $\Psi$ of our system are normalized, i.e. have unit norm, the proof of this statement is straightforward. For the goals of our study it is important to note that 
such additional relations between constraints of the system may complicate applications of the symmetry definition developed above.

\end{document}